\documentstyle[12pt]{article}
\begin{document}
\begin{center}
{\bf\large Spatial Solitons\\}
{\bf\large in Media with Delayed-Response Optical Nonlinearities}
\vspace{1cm}

{\large E.V.D}OKTOROV and {\large R.A.V}LASOV\\
{\it B.I.Stepanov Institute of Physics, F.Skaryna Ave. 70, 220072 Minsk,
Republic of Belarus}

\end{center}
\begin{tabular}{lll}
&PACS. 0340K - & Waves and wave propagation:\\
& & general mathematical aspects.\\
&PACS. 4265J - & Beam trapping, self focusing, thermal blooming,\\
& & and related effects\\
\end{tabular}

\begin{abstract}

Near-soliton scanning light-beam propagation in media with both
delayed-response
Kerr-type and thermal nonlinearities is analyzed. The delayed-response part
of the Kerr nonlinearity is shown to be competitive as compared to the
thermal nonlinearity, and relevant contributions to a distortion of the
soliton form and phase can be mutually compensated. This quasi-soliton beam
propagation regime keeps properties of the inclined self-trapped channel.

\end{abstract}

{\bf1. Introduction}
\medskip

A competition between the cubic and thermal medium nonlinearities under
different laser pulse durations seems to be firstly considered in \cite{1}.
Recently we noted \cite{2} some special features of the soliton planar
(two-dimensional) light beam propagation in a nonlinear medium with both the
promt Kerr-type nonlinearity and the inherently delayed-response thermal one.
Under condition of scanning a powerful light beam, the Kerr-type nonlinearity
was shown to be able to become dominant, while the thermal one being a
perturbation exerting an influence upon the form and phase of a Kerr soliton.
In this situation the planar beam keeps the auto-waveguide properties.

In this communication, we pay attention to some effects which occur when
taking account of the material response
of both simultaneously existing nonlinearities.
Fast scanning light beam is the case where the dynamic cubic
nonlinearity can be appreciable. It is shown that under certain condition
the delayed-response
part of the Kerr-type nonlinearity and the thermal nonlinearity
can be mutually cancelled.\\
\bigskip
\newpage
\noindent{\bf2. Formulation of the problem}
\medskip

A system of equations governing the planar light beam propagation, allowing
for both above nonlinearities and under the assumption of slight absorption,
reads \cite{3,4}

$$E_Z+{1\over u}E_t + {i\over 2k}E_{XX} = -i{k\over n_0}(\Delta n^{(K)} +
\Delta n^{(T)})E, \qquad \Delta n^{(T)}=n_T\,T,$$
$$
\tau(\Delta n^{(K)})_t + \Delta n^{(K)} = n_2|E|^2, \qquad T_t=\chi T_{XX}+
{cn_0\tilde\gamma\over2\pi\rho c_p}|E|^2.
\eqno(1)$$
Here $E$ is the beam field, $T$ is the medium temperature measured with
respect to the equilibrium one $T_0$, $\tilde\gamma$ is the linear absorption
coefficient, $\chi$ is the temperature conductivity coefficient (the heat
conductivity coefficient
 $\sigma=c_p\rho\chi$), $c_p$ is the specific heat capacity at constant
pressure, $\rho$ is the medium density, $n_0$ is the linear refraction
coefficient, $n_2$ is the Kerr constant, $n=n_o+\Delta n$, $\Delta n=
\Delta n^{(K)}+\Delta n^{(T)}$ characterizes the contributions of the
Kerr-type nonlinearity ($K$) and the thermal one ($T$), respectively, $k_0$
is the wave number, $k=n_0k_0$, $\tau$ is the relaxation time of the
Kerr-type nonlinearity, $X$ is the lateral coordinate directed along the
interface, $Z$ is the longitudinal coordinate directed into the bulk of the
nonlinear medium, $u$ is the group velocity in the first approximation of
the dispersion theory, $t$ is the running time.
There is no $Y$ dependence according to the two-dimensional model.

Let the action of a powerful radiation on the nonlinear medium be realized
via translational scanning along the $X$ axis of the planar beam, the
incidence direction being normal to the interface and coincident with the
$Z$ axis. Also, let $v$ be the scanning velocity and $D$ be a width of the
beam. Assume that the field $E$ and the temperature $T$ are dependent only
on the moving frame
coordinates $X'=X-v(t-(Z/u))$ and $Z'=Z$, being independent
on time explicitly, i.e., the stationarity condition holds in the moving
reference system. In this system we have
$$
E_{Z'}+{i\over2k}E_{X'X'}=-i{k\over n_0}(\Delta n^{(K)}+\Delta n^{(T)})E,
\qquad \Delta n^{(T)}=n_T\,T,$$
$$
\Delta n^{(K)}\approx n_2|E|^2+\tau vn_2(|E|^2)_{X'}, \qquad
-vT_{X'}={cn_0\tilde\gamma\over8\pi\rho c_p}|E|^2.
\eqno(2)$$
The expression for $\Delta n^{(K)}$ is obtained from an approximate solution
of the corresponding equation (1) under the assumption of smallness for
$\tau$ \cite{5}. Note that the small term containing the derivative of
$|E|^2$ can have another origin related to the Raman process \cite{6}.
Besides, the second order derivative in temperature conductivity equation
(1) may be neglected when $(\chi/vD)<<1$ \cite{3},which permits to represent
$T$ as $T\sim\int_{-\infty}^X|E|^2\,dX'$. Eventually, using (2),
 we obtain a
resulting equation for $E$ representable in a convenient dimensionless form:
$$
iA_z-A_{xx}-2|A|^2A=\alpha(|A|^2)_xA\pm\beta A\int_{-\infty}^x|A|^2\,dx'.
\eqno(3)$$
Here the following dimensionless quantities are introduced:
$$
x={X'\over D}, \qquad z={Z'\over{2kD^2}},
 \qquad A={1\over4}\left({c\,n_0\,g}\over
{\pi\,\sigma\, T_0}\right)^{1/2}E,$$
$$
g=16\,\pi k^3\,\sigma\,
 T_0\left({D\over n_0}\right)^2{n_2\over c}, \qquad \alpha=32\pi k^3
{{\sigma\,T_0\,D\,n_2\tau}\over{ c\,n_0^2\,g}}v,
\qquad \beta={\mu\delta\over g},
\eqno(4)$$
$$
\delta=2kD\chi\tilde\gamma\, v^{-1}, \qquad \mu=2(kD)^2{T\over n_0}|n_T|.$$
The signs $\pm$ for the last term in (3) correspond to the defocusing or
focusing effect of the thermal nonlinearity, respectively.

The coefficients $\alpha$ and $\beta$ are seen from the derivation of
the system (1) to be small ($\alpha<<1$ and $\beta<<1$), which allows Eq. (3)
to be considered as a perturbed nonlinear Schr\"odinger equation (NSE).
In the following, we shall conceive that the soliton relevant to the
non-perturbative NSE is formed at the interface. Our task is to describe
a behavior of the soliton under the action of the perturbation (3). For
definiteness, we restrict ourselves to a consideration of the defocusing
thermal nonlinearity.\\
\bigskip

\noindent{\bf3. Results.}
\medskip

In the adiabatic approximation where a distortion of the soliton form is
neglected, we seek a solution of Eq. (3) as
$$
A_s(x,z)=2\eta(z)\exp(-i\theta(z))\mbox{sech}y,
$$
where $y=2\eta(x-\kappa(z))$, $\theta=[\xi(z)/\eta(z)]y+\Delta(z)$ and
z-evolution of the parameters $\xi$, $\eta$, $\kappa$ and $\Delta$ is given
by the formulae \cite{7}:
$$
\xi_z={1\over2}\Re\int_{-\infty}^{\infty}R\,\mbox{tahn}y\,\mbox{sech}y\,
e^{i\theta}dy, \qquad \eta_z=-{1\over2}\Im\int_{-\infty}^\infty\,R\,
\mbox{sech}y\,e^{i\theta}dy,$$
$$
\kappa_z-4\xi={1\over4\eta^2}\Im\int_{-\infty}^\infty\,y\,R\,\mbox{sech}y\,
e^{i\theta}dy,
\eqno(5)$$
$$
\Delta_z-2\xi\kappa_z+4(\xi^2-\eta^2)={1\over2\eta}\Re\int_{-\infty}^\infty\,
R(1-y\,\mbox{tahn}y)\mbox{sech}y\,e^{i\theta}dy.
$$
Here the perturbation is given by
$$
R=\alpha(|A|^2)_xA+\beta A\int_{-\infty}^x|A|^2dx'.
\eqno(6)$$
Solving the equations (5) we find
$$
\xi=\xi_0+{4\over15}\gamma\,\eta_0^2\,z, \qquad \eta=\eta_0, \qquad
\kappa=4\xi_0\,z+{8\over15}\gamma\,\eta_0^2\,z^2+\kappa_0,$$
$$
\Delta=4(\xi_0^2+\eta_0^2-{1\over2}\beta\,\eta)z+{16\over15}\gamma\,\xi_0^2\,
\eta_0^2\,z^2+{4\over3}({4\over15}\gamma\,\eta_0^2)^2+\Delta_0,
\eqno(7)$$
where
$$
\gamma=16\alpha\,\eta_0^2-5\beta.
\eqno(8)$$
The soliton amplitude $2\eta$ is seen to be independent on the perturbation
while an angle of inclination of the self-trapped channel to the $z$ axis
(related to the parameter $\xi$) varies, an increase or decrease of this
angle being determined by the sign of a certain combination $\gamma$ (8)
of the parameters $\alpha$ and $\beta$. The function $\kappa(z)$ characterizes
a soliton centre location and depends on the perturbation also by means of
$\gamma$. The soliton phase $\theta(z)$ acquires a nonlinear dependence on
$z$, the latter being completely determined by perturbations nonlocal in $x$.

Finding corrections to the soliton form enables a conclusion to be drawn in
relation to keeping or non-keeping a soliton localization under a
perturbation. In the framework of the first-order approximation where a
solution of Eq. (3) is written as $A=A_s+A_1$, the correction $A_1$ is
evaluated by the formula \cite{7,2}
$$
A_1=-{1\over\pi}\Biggl(\int_{-\infty}^\infty d\lambda{\bar b(\lambda)\over{
(\lambda-\lambda_1)(\lambda-\bar\lambda_1)}}e^{-2i\lambda x}(\lambda-\xi-
i\eta\,\mbox{tahn}y)^2$$
$$
+\eta^2\,e^{-2i\theta}\mbox{sech}^2y\int_{-\infty}^\infty d\lambda{b(\lambda)
\over{(\lambda-\lambda_1)(\lambda-\bar\lambda_1)}}e^{2i\lambda x}\Biggr),
$$
where $\lambda_1=\xi+i\eta$, the bar implies complex conjugation, and
$b(\lambda)$ depends on the perturbation (6) (see \cite{2}):
$$
b(z,\,\lambda)={i\pi\over30}(\lambda-\xi)\left[12\alpha+{\gamma\over
{(\lambda-\lambda_1)(\lambda-\bar\lambda_1)}}\right]e^{i(\Delta-2\lambda
\,\kappa)}\mbox{sech}{\pi\over2}{{\lambda-\xi}\over\eta}.
$$
Restricting ourselves to finding asymptotics, we obtain
$$
x\to+\infty: \qquad A_1\to e^{-y-i\theta}\left[{16\over15}\alpha\,\eta^2+
{1\over15}\gamma\,y(y-2)\right],$$
$$
x\to-\infty: \qquad A_1\to e^{y-i\theta}\left[{16\over15}\alpha\,\eta^2-
{1\over15}\gamma\,y(y-2)\right].
$$
Thus, the soliton localization keeps in the presence of the perturbation (6)
though a slight disturbance of the initial soliton $x$-symmetry is observed.\\
\bigskip

\noindent{\bf4. Discussion}
\medskip

To illustrate the plausibility of the idea proposed, let us adduce some
typical parameters of available nonlinear condenced media and optical beams
(order-of-magnitude estimates):
$$
n_0\sim1,\qquad n_2\sim10^{-13}\ldots10^{-12}\mbox{CGSE},\qquad
\left|{{\partial n}\over{\partial T}}\right|\sim10^{-4}\ldots10^{-3}\mbox
{K}^{-1},
$$
$$
\tau\sim10^{-12}\ldots10^{-11}\mbox{s},\qquad D\sim10^{-1}\mbox{cm},
\qquad v\sim10^8\mbox{cm/s},
$$
$$
\rho c_p\sim10^7\mbox{erg/cm}^3\mbox{K},
 \qquad \tilde\gamma\sim10^{-4}\mbox{cm}^{-1},
\qquad k\sim10^5\mbox{cm}^{-1}.
$$
The magnitudes of the coefficients $\alpha$ and $\beta$ calculated with the
above parameters confirm our conjecture concerning the smallness of
$\alpha$ and $\beta$. Thus, the pertinent experimental conditions seem to be
quite realizable, including a corresponding scanning velocity value. The
last is achievable by means of up-to-date deflection techniques.

The perturbation (6) is described by two terms each of them reflects
nonlocal character of the delayed-response
nonlinearity. The non-locality of the
perturbation determined by the inertial Kerr nonlinearity is of
differential type while the non-locality due to the thermal nonlinearity
is associated with the integral type. In the moving reference frame the
above non-locality is actually similar to the spatial dispersion
influencing the soliton stability \cite{8}.

If the coefficients $\alpha$ and
$\beta$ are of different signs, the action of both perturbations leads to a
summation effect. On the other hand, if $\alpha$ and $\beta$ are of the same
sign, the nonlinear perturbations able to be compensated partially or
completely. This is seen from relevant corrections (7). A complete
compensation occurs at $\gamma=0$. Let us write down $\alpha$ and $\beta$
isolating the explicit dependence on the scanning velocity: $\alpha=
\alpha^*v$ and $\beta=\beta^*v^{-1}$ (expressions for $\alpha^*$ and
$\beta^*$ can be readily obtained from (4)). Then we get from the condition
$\gamma=0$:
$$
v^2={5\over16}{\beta^*\over{\alpha^*\,\eta^2}}.
\eqno(9)$$
Hence, a feasibility of nonlinear distortion compensation is due to
difference in $v$ dependence of both nonlinearities. Namely, the influence
of the Kerr-type nonlinearity increases with increasing $v$, while that of
the thermal nonlinearity decreases. Note that the relation (9) resembles
the resonance condition for the oscillating contour.

Thus, a compensation of distortions caused by the delayed-response part
of the cubic nonlinearity and
defocusing thermal nonlinearities leads, in the adiabatic approximation, to
the "pure" soliton having the envelope of sech-type and the phase linear in
$z$. In this case, there remains a linear correction due to the thermal
nonlinearity. It is worth to point out that one can also use tilted pulses
\cite{9} which resemble translationally scanned beams.\\

\bigskip
\begin{centerline}
{* * *}
\end{centerline}

The authors are grateful to one of the referees for constructive suggestions.
This work was supported in part by the Fundamental Research Foundation of
Belarus under Contract No. $\Phi$-17/205.

\end{document}